\documentstyle[12pt,epsf]{article}

\oddsidemargin--0mm
\newcommand{\be}{\begin{equation}}
\newcommand{\ee}{\end{equation}}
\newcommand{\bea}{\begin{eqnarray}}
\newcommand{\eea}{\end{eqnarray}}

\def\f{\frac}
\def\p{\partial}
\def\l{\lambda}
\def\t{\tau}
\def\s{\sigma}

\def\le{\left}
\def\ri{\right}
\def\nn{\nonumber}
\def\nb{\nabla}

\begin{document}

\begin{titlepage}

\vspace*{1cm}
\begin{center}
{\Large\bf
Nambu--Goto string with the Gauss--Bonnet term \\
 \vskip 3mm
 and point--like masses at the ends}

\vskip 1cm
{\large{\bf Leszek Hadasz$^{\dag}$} and {\bf Tomasz R\'og$^{\ddag}$}}

\vskip 1cm
Jagellonian University,  \\
M.Smoluchowski Institute of Physics \\
Reymonta 4, 30--059 Krak\'ow, Poland
\end{center}

\vskip 1cm
\begin{abstract}
In the present paper we investigate classical dynamics of the
Nambu--Goto string with Gauss--Bonnet term in the
action and point--like masses at the ends in the context
of effective QCD string. The configuration
of rigidly rotating string is studied and its
application to phenomenological description of meson
spectroscopy is discussed.
\end{abstract}
\vspace{\fill}

\noindent
\begin{flushleft}
TPJU -- 15/96 \\
July 1996 
\end{flushleft}

\vspace{1.4cm}

\noindent
\underline{\hspace*{16cm}}

\noindent
$ ^{\dag}$E--mail: hadasz@thp1.if.uj.edu.pl

\noindent
$ ^{\ddag}$E--mail: ufrog@thp1.if.uj.edu.pl

\end{titlepage}
The existence of some, at least approximate string representation of QCD,
although not rigorously proved, seems to be plausible and useful
\cite{pol}. It is supported by the nature
of the $1/N_c$ expansion \cite{intr1}, the success of dual models in
description of Regge phenomenology, area confinement law found
in the strong coupling lattice expansion \cite{intr3} or the existence
of flux--line solutions in confining gauge theories \cite{intr4,nkap3} and
the analytical results concerning two--dimensional QCD \cite{intr5}.
More arguments can be found in review \cite{intr6}.

It is an obvious statement that the role of boundary conditions
imposed on the fields can be crucial for the predictions of a
theory. The same is also
true in  string theory --- for example, the spectra of states
in the case of closed Nambu--Goto \cite{ng} string theory (with the periodic
boundary conditions) and in the case of the open Nambu--Goto string with free
ends, satisfying the Neumann boundary conditions, are essentially different.
Moreover, in the effective string
description of QCD quark's internal degrees of freedom
(colour and electric charges, spin or mass) enter the
theory through the boundary conditions, \cite{leh1}.

Motivated by such considerations we are led to study the impact
of a  modifications in the boundary conditions on the string's
dynamics. We shall consider in the present paper the simplest,
Nambu--Goto string model, being also the most natural
zeroth order approximation to the QCD string.

If we confine ourselves to the string action which is a total derivative
(and consequently affects only the boundary conditions for the string),
depends only on the first and second order derivatives of the vectors
$X^\mu(\tau,\sigma)$ defining the immersion of the string in the
four dimensional, Minkowski space--time, and satisfying the
natural requirements of being Poincare' as
well as reparametrization (with respect to $\tau$
and $\sigma$) invariant, then we are led \cite{paw}
to the unique possibility,
\be
S_{\rm b} = -\frac{\alpha}{2}S_{\rm GB} - \beta S_{\rm Ch}.
\ee
Here $\alpha$ and $\beta$ are dimensionless parameters $(\alpha > 0).$
$S_{\rm GB}$ and $S_{\rm Ch}$ are pseudoeuclidean Gauss--Bonnet and
Chern terms which, upon Wick rotation to the Euclidean space, are related
to the Euler characteristic of the surface and to the number of its
self--intersections, respectively.

If one further wishes to take into account the non--zero masses of quarks,
then it is naturally to add to the action the contributions from the
massive point--like particles attached to the string ends,
\be
S_{\rm p} = -m_1L_1 - m_2L_2,
\ee
where $L_1$ and $L_2$ are invariant lengths of the trajectories of the
string ends.

More explicitly, the action of the
Nambu--Goto string with  modified boundary conditions and point--like
masses at the ends is of the form:
\be
\label{r1}
S = \int_{\t_1}^{\t_2}\!\!d\t\int_{\s_1(\t)}^{\s_2(\t)}\!\!d\s
\sqrt{-g}{\cal L} -
\sum_{i=1}^2\;m_i\int_{\t_1}^{\t_2}\!dt\;\sqrt{(d_tX)^2}|_{\s = \s_i(\t)},
\ee
with
\be
{\cal L} = -\gamma - \f12\alpha R -\beta N.
\ee
Here $\gamma$ is the string tension,
$t$ parametrizes points along the trajectories of the string ends and
$g = \det g_{ab}$ with
$g_{ab} = \p_aX^\mu\p_bX_\mu$ being the induced metrics. The curvature scalar
$R$ and the integrand of the Chern term $N$ are
defined according to the formulae
\bea
R & = & \le(g^{ab}g^{cd} - g^{ad}g^{bc}\ri)\nb_a\nb_bX^\mu\;\nb_c\nb_dX_\mu, \\
&& \nn \\
N & = & -\f{1}{2\sqrt{-g}}g^{ac}\epsilon^{bd}\tilde t^{\mu\nu}
           \nb_a\nb_bX^\mu\;\nb_c\nb_dX_\mu,
\eea
where $\nb_a$ is the covariant (with respect to the metric $g_{ab}$) derivative
and
$$
\tilde t^{\mu\nu} = \f{1}{\sqrt{-g}}\epsilon^{\mu\nu\rho\l}
     \p_\t X_\rho\p_\s X_\l.
$$

In the following we shall assume that the space--like parameter
$\s$ varies in the interval $[0,\pi].$
It is also convenient to fix the worldsheet parametrization by imposing
the conditions \cite{bnest}:
\bea
\label{r2}
(\dot X \pm X')^2 & = & 0, \\
&& \nn \\
(\ddot X \pm \dot X{}')^2 & = & -\f14q^2,
\eea
where the dot and the prime mean differentiation with respect to $\t$ and
$\s,$ respectively, and $q$ is an arbitrary parameter with the dimension
of mass.
It can be shown then (for details see \cite{paw,bnest,paw2}), that
every solution of the string
equations of motion and boundary conditions, following from the action
(\ref{r1}), corresponds to the
solution of the complex Liouville equation \cite{liouv}:
\be
\label{liouv}
\ddot\Phi - \Phi'' = 2q^2e^{\Phi},
\ee
supplemented with the boundary conditions:
\be
 \label{bc}
 \le\{
 \begin{array}{lcll}
 \gamma - \alpha q^2e^{2\Re\Phi} & = &
    (-1)^im_i\f{\p}{\p\s}\le(e^{\Re\Phi/2}\ri), &  \\
 &&&  \\
 C\f{\p}{\p\t}\Re\Phi & = & 0, & \\
 &&& \hspace*{1cm} \mbox{for} ~\s = 0,\pi, \\
 C\cos\le(\Im\Phi/2\ri) & = & \beta, & \\
 &&& \\
 C^2\f{\p}{\p\s}\Im\Phi & = &
   2\beta(-1)^im_ie^{-\Re\Phi/2}, &
 \end{array}
 \ri.  
\ee
where $C = \sqrt{\alpha^2+\beta^2}.$
This correspondence is explicitly established through the relations:
\bea
\label{con1}
 e^{\Phi} & = & -\f{1}{q^2}\f{F'_L(\t+\s)F'_R(\t-\s)}{
     \sin^2\le[\f{F_L(\t+\s)-F_R(\t-\s)}{2}\ri]},  \\
 && \nn \\
\label{con2}
 X^\mu(\t,\s) & = &  X^\mu_L(\t+\s) + X^\mu_R(\t-\s),  \\
 && \nn \\
\label{con3}
 \f{\p}{\p\t}X^\mu_{L,R} & = &
 \f{q}{2|F'_{L,R}|}\le(\cosh\Im F_{L,R},\;\cos\Re F_{L,R},\;\sin\Re F_{L,R},\;
               \sinh\Im F_{L,R}\ri),
\eea
where $F_{L,R}$ are arbitrary complex functions which give single valued
$\Phi$ satisfying the boundary conditions (\ref{bc}).

We shall also need the formulae for the string's momentum
and angular momentum. The contributions to these quantities from the
particles located at the string ends are standard so there is no
need to write
them down here explicitly while the contributions from the surface
terms in (\ref{r1}) can be obtained from the general formulae:
\bea
\label{r3}
P_{\mu} & = & \int_0^{\pi}\!\!d\s \;\sqrt{-g}\;\Pi_{\mu}^0 -
\left[\sqrt{-g}\f{\p{\cal L}}{\p(\nb_0\nb_1X^{\mu})}\right]_0^{\pi},
\nn \\
&& \\
M_{\mu\nu} & = & \int_0^{\pi}\!\!d\s \;\sqrt{-g}\le\{X_{[\mu}\Pi^0_{\nu]} -
X_{[\mu,a}\f{\p{\cal L}}{\p(\nb_a\nb_0X^{\nu]})}\ri\} -
\left[\sqrt{-g}X_{[\mu}\f{\p{\cal L}}{\p(\nb_0\nb_1X^{\nu]})}
\right]_0^{\pi}, \nn
\eea
where
\be
\label{r5}
\Pi^a_{\mu} = -{\cal L}\nb^aX_{\mu} - \f{\p{\cal L}}{\p X^{\mu}_{,a}} +
      2\f{\p{\cal L}}{\p g^{bc}}g^{ab}\nb^cX_\mu +
      \nb_b\left[\f{\p{\cal L}}{\p(\nb_a\nb_bX^{\mu})}\right].
\ee
The formulae (\ref{r3},\ref{r5}) can be applied in the case of general
lagrangian ${\cal L}$ being the scalar function of the vector $X^\mu(\t,\s)$
and its covariant derivatives up to the second order and turn out to be
quite useful in the practical calculations involving the action (\ref{r1}).

In the following we shall consider only the special case $\beta = 0$ of
the general model described by the action (\ref{r1}).
The reason for this is that the generic classical solutions of the
model with the Chern term correspond to the
self--accelerating strings which are hard to interpret from the point of
view of the effective QCD string theory and, moreover,
this solutions exist only if some
(different for different solutions) relations between the parameters of the
model ($\gamma,\alpha,\beta$ and $m_i$--s) are satisfied.

A distinguished class of solutions of the Liouville equation
(\ref{liouv}) is composed of static, i.e. $\t$--independent
fields. They are of the form
\be
\label{sol1}
e^\Phi = -\f{\l^2}{q^2}\f{1}{\cos^2\le(\l\s-d\ri)},
\ee
where $\l$ and $d$ are arbitrary, complex constants.

>From the Eqs. (\ref{sol1}) and (\ref{con1}) we get
\bea
\label{r6}
F_L & = & \l(\t+\s) - d, \nn \\
&& \\
F_R & = & \l(\t-\s) + d - \pi. \nn
\eea

To check whether this is actually a solution we have to take into account the
boundary conditions (\ref{bc}). They imply that $\l$ and $d$ are real
numbers satisfying the equations:
\bea
\label{bc2}
\f{\gamma q}{\l^2}\cos^4 d - m_1\sin d\,\cos^2 d -
\f{\alpha\l^2}{q} & = & 0, \nn \\
&& \\
\f{\gamma q}{\l^2}\cos^4(\pi\l-d) - m_2\sin(\pi\l-d)\,\cos^2(\pi\l-d) -
\f{\alpha\l^2}{q} & = & 0. \nn
\eea
Equations (\ref{bc2}) can  be easily solved numerically for any specific
values of the parameters $\gamma,\alpha$ and $m_i,$ giving a
family of solutions parametrized by the constant $q.$

In the language of string variables the static Liouville field
(\ref{sol1}) describes
the string which rotates rigidly in the $X^3 = 0$ plane,
\be
\label{rod}
X^\mu(\t,\s) = \f{q}{\l^2}(\l\t,\;\cos\l\t\,\sin(\l\s-d),\;
                            \sin\l\t\,\sin(\l\s-d),\;0).
\ee
Using the formulae
(\ref{r3},\ref{r5}) we can calculate its energy and the third component
of the angular momentum,
\bea
\label{ene}
E & = & \f{\pi q\gamma}{\l}\le[1+\f{\sin\pi\l\,\cos(\pi\l-2d)}{\pi\l}\ri]
        + m_1\cos d + m_2\cos(\pi\l-d),  \\
&& 	\nn \\
\label{ang}
J & = & \f{\pi q^2\gamma}{2\l^3}\le[1 +
    2\f{\sin\pi\l\,\cos(\pi\l-2d)}{\pi\l} +
     \f{\sin2\pi\l\,\cos(2\pi\l-4d)}{2\pi\l}\ri] - \nn \\
&& \\
&& \hspace*{1cm} -\f{m_1q}{\l^2}\sin^2d\,\cos d -
                  \f{m_2q}{\l^2}\sin^2(\pi\l-d)\,\cos(\pi\l-d), \nn
\eea
with the total spatial momentum and the other components of the angular
momentum being zero. The values of $\l = \l(q)$ and $d = d(q)$ which appear
in (\ref{ene}) and (\ref{ang}) are determined from the Eqs. (\ref{bc2}).

String which rigidly rotates in a plane constitutes a configuration
with maximal angular momentum at a given energy. Such configurations
compose a classical, leading  Regge trajectory.
For large values of $q$ one can determine using (\ref{bc2}), (\ref{ene})
and (\ref{ang}) the asymptotic form of the relationship between the angular
momentum and energy squared to be
\be
J = \f{E^2}{2\pi\gamma} + \f{1}{3\gamma}\sqrt{\f{2}{\pi}}
 \le[\le(\sqrt{m_1^2+4\alpha\gamma}-2m_1\ri)
   \sqrt{m_1+\sqrt{m_1^2+4\alpha\gamma}} + \le(m_1 \to m_2\ri)\ri]E^{\f12} +
   {\cal O}(E^0).
\ee
This formula clearly shows that the pertinent Regge trajectory in the
discussed model can be raised or lowered with respect to the
Regge trajectory in the ``pure'' Nambu--Goto model,
$$
J_{\rm NG} = \f{1}{2\pi\gamma}E^2_{\rm NG},
$$
depending on the ratio of the point--like masses $m_i$ squared to the
product $\alpha\gamma.$

Although the discussed model is rather simple it is amusing
to compare its predictions with the data concerning the spectrum of
mesons. One finds that the parameters of the model can be
nicely fitted to the energy and angular momentum of the mesonic states
lying on the leading Regge trajectories both for the heavy--heavy
and heavy--light systems (which is not surprising, as the model of
Nambu--Goto string with point--like masses and {\em without} Gauss--Bonnet
term already reproduces the data quite well, \cite{barb}) and for
the light--light
systems, where the model without Gauss--Bonnet term breaks down.
The example is plotted on the Fig.1.

\centerline{\epsfbox{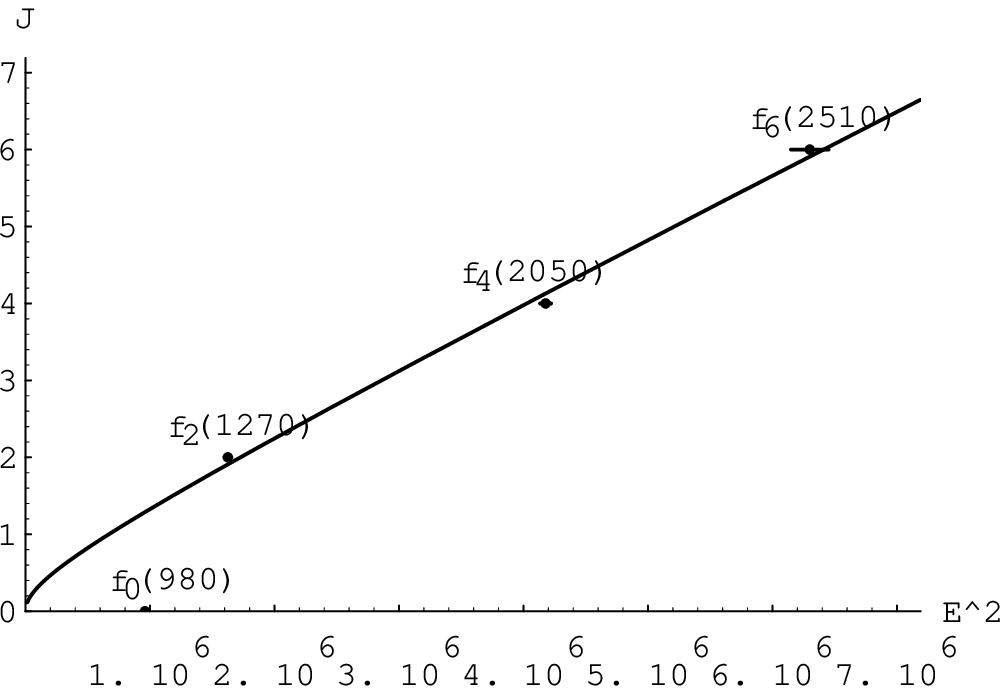}}

\begin{center}
{\small Fig.1 Regge trajectory of the model fitted to the data
from the meson $f$ family. The values of the parameters are $m_1 = m_2 =0,$
$\gamma = 1.96\times10^5 (MeV)^2$ and $\alpha = 0.19.$}
\end{center}

\vskip 5mm

Let us note a few facts:
\begin{itemize}
\item
the (approximately) linear asymptotes of the
classical Regge trajectories obtained in the discussed model
have  non--zero intersection with the angular momentum axis, while
in the ``pure'' Nambu--Goto case the quantum corrections are
needed to achieve this,
\item
the ``dother'' Regge trajectories can be described by considering
oscillations of the string around the ``rotating rigid rod''
configuration (their explicit form can be obtained in a way
analogous to the one used in \cite{lpaw}),
\item
the fitted values of the string tension $\gamma$ and the parameter
$\alpha$ are approximately independent of the meson family
(i.e. are approximately the same for the heavy--heavy, heavy--light
and light--light systems),
\item
for the light-light systems (excluding
the lightest meson in each family --- we shall comment on this below)
we obtain the quark masses equal approximately to zero, i.e. the
point--like masses in the model seems to be connected with the
light, current quarks rather then with the heavy constituent
quarks of the non--relativistic quark model.
\end{itemize}
The last point is probably less surprising if one takes into account
that addition to the Nambu--Goto string of the Gauss--Bonnet term
already results in a non--uniform distribution of the energy density,
with the maxima at the string ends, \cite{lpaw}.

The model seems to have problems with the light mesons of zero spin.
Indeed, as illustrated on the Fig.1, for the zero point--like
masses the classical solution with zero angular momentum has
vanishing energy. But the region of slowly rotating string is
precisely the one where the quantum corrections become important.
In the paper \cite{le} the first quantum correction (the
Casimir energy, \cite{cas}) to the classical energy of the rigidly
rotating
string configuration in the model of Nambu--Goto string with
the Gauss--Bonnet term has been computed using the semiclassical
approximation.
This correction grows indefinitely for the strings rotating
with vanishing frequency, invalidating there the classical picture.
The inclusion of point--like masses at the string ends does not
qualitatively change this result and, similarly as in the model
without point--like masses, the lightest string solution has
non--zero energy. It is then naturally to expect that this
quasiclasically corrected, lightest string configuration
will be a correct candidate for the scalar meson in the
discussed effective model.

\vskip 3mm
\noindent
{\large\bf Acknowledgments}

The authors would like to thank dr P. W\c{e}grzyn for inspiration
and numerous helpful discussions. One of us (L.H.) was supported
by grant KBN 2 P03B 045 10 and by Foundation for Polish Science
scholarship.


\begin{thebibliography}{99}
 \bibitem{pol}
  A.M.Polyakov, {\em Gauge fields and strings,}  Harwood
   Academic Press, 1987.
\bibitem{intr1}
  G.'t Hooft, Nucl.Phys {\bf B72} (1974) 461.
\bibitem{intr3}
  K.Wilson, Phys.Rev {\bf D8} (1974) 2445.
\bibitem{intr4}
  H.B.Nielsen and P.Olesen, Nucl.Phys. {\bf B61} (1973) 45.
\bibitem{nkap3}
  M.Baker, J.S.Ball and F.Zachariasen, Phys.Rev. {\bf D41} (1990) 2612.
\bibitem{intr5}
   David J.Gross, Nucl.Phys. {\bf B400} (1993) 161.
   David J.Gross and Washinghton Taylor IV, Nucl.Phys. {\bf B400} (1993) 181.
   David J.Gross and Washinghton Taylor IV, Nucl.Phys. {\bf B403} (1993) 395.
 \bibitem{intr6}
  J.Polchinski, {\em Strings and QCD?,} preprint UTTG-92-16 and
  hep-th/9210045.
\bibitem{ng}
  Y.Nambu, Lectures on the Copenhagen Summer Symposium (1970), unpublished,
   O.Hara, Prog.Theor.Phys. {\bf 46} (1971) 1549,
   T.Goto, Prog.Theor.Phys. {\bf 46} (1971) 1560,
   L.N.Chang and J.Mansouri, Phys.Rev. {\bf D5} (1972) 2535,
  J.Mansouri and Y.Nambu, Phys.Lett. {\bf 39B} (1072) 375.
\bibitem{leh1}
   L.Hadasz, Phys.Lett. {\bf B324} (1994) 36,
   L.Hadasz, Acta Phys.Pol. {\bf B25} (1994) 1419.
\bibitem{paw}
   P.W\c{e}grzyn, Phys.Rev. {\bf D50} (1994) 2769.
\bibitem{bnest}
  B.M.Barbashov, V.V.Nesterenko {\em Introduction to the relativistic
  string theory,} World Sci., Singapore 1990.
\bibitem{paw2}
   P.W\c{e}grzyn, {\em Strings with interacting ends,} preprint TPJU--23/94,
   Mod.Phys.Lett. {\bf A}, in press;
   J.Karkowski, Z.\'Swierczy\'nski and P.W\c{egrzyn},
   {\em Nambu--Goto string action with Gauss-Bonnet term,}
   preprint TPJU--24/94, Mod.Phys.Lett. {\bf A}, in press.
\bibitem{liouv}
   J.Liouville, Math.Phys.P.Appl. {\bf 18} (1853) 71,
   A.R.Forshyt, {\em Theory of Differential Equations,} Part4, vol. 6,
   New York, Dover, 1959,
   G.P.Jorjadze, A.K.Pogrebkov, M.C.Polivanov, Doklady Akad. Nauk SSSR
   {\bf 243} (1978) 318.
\bibitem{barb}
   B.M.Barbashov, {\em Classical dynamics of Rotating Relativistic String
   with Massive Ends: the Regge Trajectories and Quark Masses,}
   preprint JINR--E2--94--444, submitted to Nucl.Phys. {\bf B}.
\bibitem{lpaw}
   L.Hadasz and P.W\c{egrzyn}, Phys.Rev. {\bf D51} (1995) 2891.
\bibitem{le}
   L.Hadasz, {\em Ground state enrgy of the Nambu--Goto string with
   modified boundary conditions,}
   preprint TPJU 14/96, submitted to Mod.Phys.Lett. {\bf A}.
\bibitem{cas}
  H.B.G.Casimir, Proc.Kon.Nederl.Akad.Wetenschap {\bf 51} (1948) 739.
\end{thebibliography}
\end{document}